\documentclass[apj]{emulateapj}

\begin{document}

\title{Environment and the cosmic evolution of star formation}

\author{Ravi K. Sheth,\altaffilmark{1} Raul Jimenez,\altaffilmark{1,2} 
        Ben Panter\altaffilmark{3}, and Alan F. Heavens\altaffilmark{4}}

\affil{}

\altaffiltext{1}{Dept. of Physics and Astronomy, University of Pennsylvania, 
                 209 South 33rd St, Philadelphia, PA 19104, U.S.A.; 
                 shethrk,raulj@physics.upenn.edu}
                 
\altaffiltext{2}{The Observatories of the Carnegie Institution, 
                 813 Santa Barbara St., Pasadena, CA 91101, U.S.A.}

\altaffiltext{3}{Max Planck Institut f\"ur Astrophysik, 
                 Karl Schwarzschild Str. 1,
                 Garching bei M\"unchen, D-85748, 
                 Germany; bdp@mpa-garching.mpg.de}

\altaffiltext{4}{Institute for Astronomy, University of Edinburgh, 
                 Blackford Hill, Edinburgh, EH9-3HJ, Scotland; afh@roe.ac.uk}

\begin{abstract}
We present a mark correlation analysis of the galaxies in the 
Sloan Digital Sky Survey using weights provided by MOPED.  
The large size of the sample permits statistically significant 
statements about how galaxies with different metallicities and 
star formation histories are spatially correlated.  
Massive objects formed a larger fraction of their stars at higher 
redshifts and over shorter timescales than did less massive objects 
(sometimes called down-sizing).  We find that those 
galaxies which dominated the cosmic star formation at 
$z \approx 3$ are predominantly in clusters today, whereas 
galaxies which dominate the star formation at $z \approx 0$ inhabit 
substantially lower mass objects in less dense regions today.
Hence, our results indicate that star formation and chemical 
enrichment occured first in the denser regions of the Universe, 
and moved to less dense regions at later times.  
\end{abstract}



\section{Introduction}

There has been substantial recent progress in the development of 
methods which determine the star formation and chemical composition 
histories of galaxies from the integrated spectra of their stellar 
populations.  
The traditional approach has been to determine the instantaneous star 
formation rate or metallicity from certain features in the spectrum 
of a galaxy.  However, several recent algorithms 
\citep{HJL00, Sodre05, Mathis06, Ocvirk06}, have been developed 
which use the entire spectrum to infer the entire star formation 
history and the evolution of the chemical composition of the object, 
rather than simply the instantaneous values of these quantities.  
One such method, MOPED \citep{HJL00}, has been used to determine the 
star formation histories and metallicities of galaxies drawn from 
the Sloan Digital Sky Survey 
\citep{PHJ03,HPJD04,PHJ04,JPHV05,PJHC06a,PJHC06b}.

There has also been significant progress in quantifying how 
galaxies are distributed on large scales, and using this to 
constrain cosmological parameters \citep{2dFPk05, SDSSPk05}.  
In such analyses of galaxy clustering, it is common to treat 
galaxies as points, ignoring the fact that galaxies have different 
luminosities, colors, masses, star formation histories, metallicities, 
etc.  However, as a result of improvements in detector technology, 
and in the algorithms such as MOPED with which the new data is 
analyzed, many such galaxy attributes are now sufficiently reliably 
measured that one can use them as weights when studying the 
clustering of galaxies.  Thus, one can now study the clustering of 
luminosity, color, star formation rate, etc.  Mark statistics 
\citep{stoyan2,BK00} provide a useful framework for describing 
attribute-weighted clustering.  Moreover, they provide sensitive 
probes of how the properties of galaxies correlate with their 
environments \citep{sheth05}.  In this respect, mark statistics 
provide a useful link between the large-scale structures which 
galaxies trace, and the properties of those galaxies.  They have 
recently been used to measure the clustering of luminosity and 
color in the SDSS \citep{SSCS06}.

In this Letter, we use mark correlations to demonstrate the wealth 
of information which the MOPED-determined histories provide.  
We show that a mark correlation analysis, using MOPED attributes 
as marks, allows one to study how the star formation histories of 
objects are correlated with their present-day environments.  
Our main finding is that the majority of close ($< 2$ Mpc) galaxy 
pairs today are made up of objects which formed the largest fraction 
of their stars at $z \approx 3$, 
whereas those which formed stars at an above average rate within 
the last Gyr or so tend to be in less clustered environments today.  
In hierarchical models, overdense regions today were overdense in the 
past.  Thus, our results indicate that star formation at high-redshift 
occured in dense regions, whereas it only occurs in substantially less 
dense regions today.  A similar study of the metallicity shows 
that the close pairs which formed a larger than average fraction 
of their stars at $z\approx 3$ also have above average 
metallicities, but that there is no correlation between 
metallicity and environment for stars formed more recently.  
Where necessary, we assume a flat $\Lambda$CDM model with 
$\Omega_0=0.27$ and $H_0=71$~km~s$^{-1}$Mpc$^{-1}$ \citep{WMAP06}.

\section{Measurements in the SDSS}\label{sample}
As described in \citet{PJHC06a,PJHC06b}, the MOPED algorithm has 
been used to extract star formation and metallicity histories of 
a magnitude limited sample ($15.0 \le m_r \le 17.77$) of about 
300,000 galaxies drawn from the Third Data Release of the 
Sloan Digital Sky Survey (SDSS DR3; \citet{SDSS-DR3}).   
We measured mark correlations, using MOPED attributes as the marks, 
in two smaller volume-limited catalogs extracted from the SDSS DR3.  
To facilitate future halo-model based interpretations of our 
measurements, these catalogs were chosen to approximately correspond 
to those studied by \citet{Zehavi05} and \citet{SSCS06}:
a brighter catalog which spans $M_r<-21.5$ for which $0.02 < z < 0.135$, 
and one which includes fainter objects, $M_r<-20$, and so spans 
a smaller redshift range: $0.02 < z < 0.071$ 
(the associated apparent magnitude limit is conservative: $m_r \le 17.5$).


\begin{figure}
 \includegraphics[width=0.9\columnwidth]{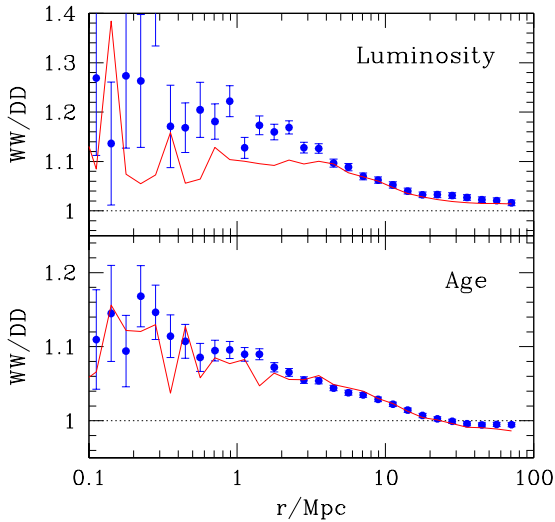}
 \caption{Mark correlations in the two volume limited catalogs 
         described in the text; symbols with error bars show results 
         for the fainter sample, and lines without error bars are for 
         the more luminous sample.     
         Close pairs tend to be more luminous (top panel) and to 
         host older stellar populations (bottom) than average.} 
 \label{fig:xiLumAge}
\end{figure}

Figure~\ref{fig:xiLumAge} shows the mark correlations in the two 
catalogs when $r-$band luminosity (top) and the MOPED-inferred 
mass-weighted age (bottom) are the marks (symbols with error bars 
show results for the fainter sample, and lines without error bars 
are for the brighter sample).  
So that the weights are dimensionless, the weight of each galaxy 
is normalized by the mean value for the population.  The notation 
WW/DD indicates that the mark statistic is the ratio of weighted 
pair counts to unweighted pair counts.  (In this notation, the 
traditional unweighted correlation function would be DD/RR, 
where RR is the number of unweighted pair counts in a random 
distribution.)  
Symbols with error bars show results for the fainter catalog, and 
lines without error bars are for the more luminous sample---we use 
a similar convention in all the figures which follow.  
Error bars were estimated using the analytic expressions given in 
\citet{SCS06}, which \citet{SSCS06} have shown are similar to those 
from a jackknife analysis; they are similar for the two samples.   
The results of Skibba et al. also show that, while using redshift- 
rather than real-space distances tends to make WW/DD closer to 
unity on small scales, this is not a severe effect.  

The top panel shows that close pairs of galaxies tend to be more 
luminous than average, consistent with previous mark correlation 
analyses of SDSS galaxies \citep{SSCS06}.  Halo-model based 
analyses of the clustering of SDSS galaxies indicate that, on scales 
smaller than 1~Mpc, the pair counts are dominated by galaxies in 
massive haloes \citep{Zehavi05}.  Thus, the top panel shows that 
galaxies in massive haloes are over-luminous.  

The bottom panel shows that, in addition to being more luminous 
than average, close pairs tend to have older than average stellar 
populations.  This suggests that the most massive halos host the 
oldest stellar populations.  
The scale dependence in the bottom panel is also qualitatively 
similar to that seen in semi-analytic galaxy formation models 
\citep{sheth05, SCS06}.  In the models, this happens because close 
pairs are dominated by galaxies in clusters, and cluster galaxies 
host the oldest stars.  This is consistent with the halo-model 
interpretation mentioned above.  

\begin{figure}
 \vspace{-2.3cm}
 \includegraphics[width=\columnwidth]{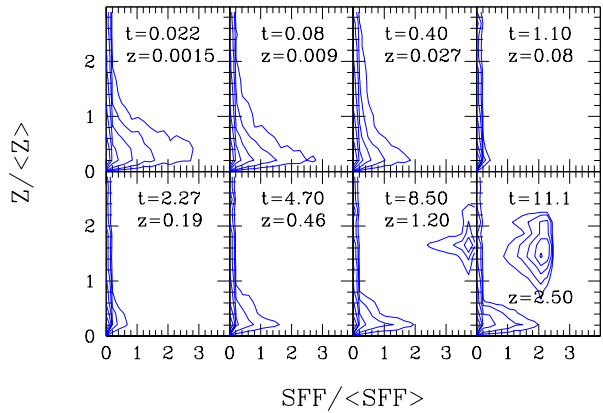}
 \caption{Joint distribution of star-formation fraction and 
          metallicity in the volume limited catalog with $M_r<-20$; 
          Different panels show results for different lookback 
          times, $t$ in Gyrs, and associated redshifts, $z$.  
          In all panels, the marks have been normalized by the 
          mean mark in the bin.  }
 \label{fig:sffMet}
\end{figure}

\begin{figure*}
 \centering
 \vspace{-5cm}
 \includegraphics[width=1.9\columnwidth]{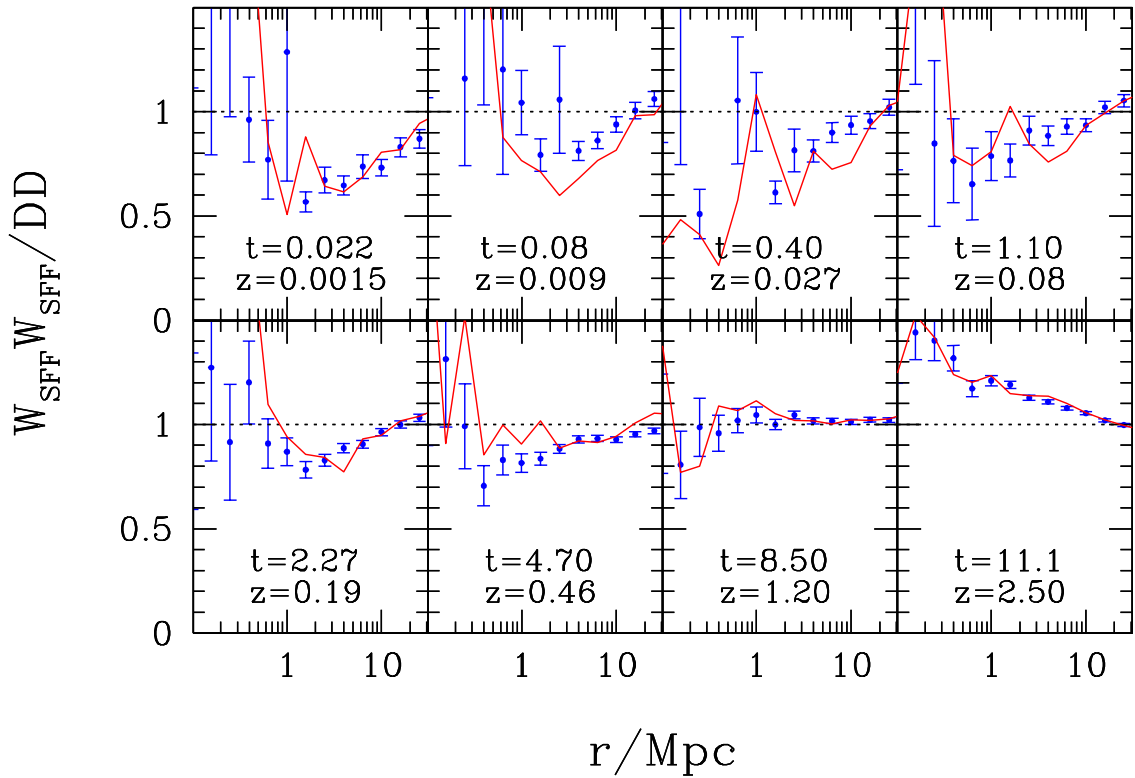}
 \caption{Mark correlation functions, with star-formation fraction 
          as the mark, in the two volume limited catalogs described 
          in the text.  Symbols with error bars show results for the 
          fainter sample; lines without error bars represent the 
          more luminous sample.  
          Panels show results for different lookback times, 
          $t$ in Gyrs, with associated formation redshifts, $z$.  
          In both catalogs:  
          close pairs today had higher than average star formation 
          fractions at $z=2.5$, average star formation fractions at $z=1$, 
          and lower than average star formation fractions more recently.  
          The anti-correlation between star formation fraction and 
          environment persists up to lookback times of 5~Gyrs.}
 \label{fig:xiSFF}
\end{figure*}

Figure~\ref{fig:xiLumAge} uses marks which are the result of 
integrating over the entire star formation history of each object.  
One of the great virtues of the MOPED analysis is that it returns 
not just the mass-weighted age at the present time, but an estimate 
of the entire star formation history of an object.  Thus, for 
each object, we have constructed estimates of the fraction of the 
current stellar mass which formed in each of eight bins in lookback 
time.  

Figure~\ref{fig:sffMet} shows the joint distribution of star 
formation fraction and metallicity in the fainter catalog, for 
each of the lookback time bins.  In each panel, the marks have 
been normalized by the mean value in the bin.  For instance, at 
the two largest lookback times, 
$(\langle {\rm SFF}\rangle, \langle {\rm Z/Z}_\odot\rangle) 
  = (0.25,0.65)$ and $(0.43,0.77)$.  These numbers are 
$(0.25,0.70)$ and $(0.55,0.89)$ in the more luminous catalog.  
The differences indicate that massive objects formed a larger 
fraction of their stars at higher redshifts, on a shorter 
timescale, than lower mass objects---a point made by \citet{HPJD04}.

In all panels there is a population of objects which have small 
star formation fractions but a large range of metallicities, and 
a population which has small metallicities but a tail which extends 
to large star formation fractions.  
However, at large lookback times, there is an additional population 
which has above average metallicity {\em and} star formation fraction.  
For the two largest lookback times, this population comprises 
24\% and 39\% of the galaxies in the fainter catalog 
(26\% and 51\% in the brighter catalog).  Where are these objects now?  

Figure~\ref{fig:xiSFF} shows a mark correlation analysis of the 
star formation fraction at these eight lookback times.  The bottom right 
panel shows that close pairs today had larger than average star 
formation fractions 11 Gyrs ago.  Comparison with the other panels 
indicates that these pairs had average star formation fractions 
at redshifts of order unity, and smaller than average star 
formation fractions more recently.  
Thus, our analysis provides graphic evidence that the 
objects which underwent vigorous star formation at the highest 
redshifts are currently in clusters, where the current star formation 
rate is smaller than average.  

\begin{figure*}
 \centering
 \vspace{-5cm}
 \includegraphics[width=1.9\columnwidth]{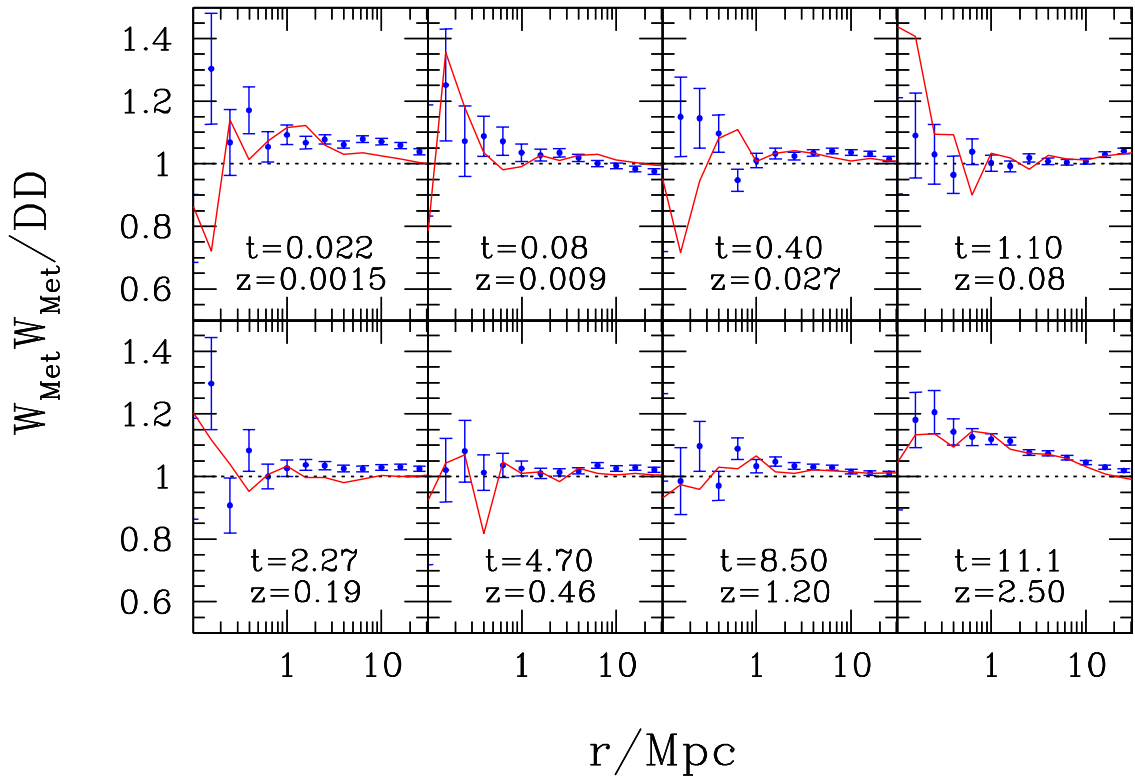}
 \caption{Mark correlation functions with metallicity as 
          the mark in the two volume limited catalogs described in 
          the text.  Symbols with error bars show results for the 
          fainter sample, and lines without error bars are for 
          the more luminous sample.  Panels show results for different 
          lookback times.  
          Close pairs which formed their stars 11~Gyrs ago tend to 
          have metallicities which are above average for that 
          epoch (bottom right).  
          At smaller lookback times there is little 
          correlation between metallicity and present day 
          environment except very recently (top left) where most 
          of the enrichment is taking place in galaxy groups.} 
 \label{fig:xiMet}
\end{figure*}

Figure~\ref{fig:xiMet} shows a similar analysis when metallicity, 
$Z/Z_\odot$, is the mark.  
The bottom right panel shows that the close pairs which had above 
average star formation fractions at $z=2.5$ also tend to have above 
average metallicities.  There are no clear correlations with 
environment in the other panels, except, possibly, at the smallest 
lookback times (top left), where there is an enhancement on scales 
where the pair counts are dominated by galaxy groups. 
Thus, our measurements indicate that the stellar populations of the 
most massive halos are old and metal rich.  The population of objects 
which had above average star formation fractions and metallicities 
at large lookback times (c.f. Figure~\ref{fig:sffMet}) are today in 
clusters.  

\section{Conclusions}

A mark correlation analysis of SDSS galaxies using MOPED-derived 
ages, metallicities and star formation histories shows that close 
pairs tend to host stellar populations which are older than average 
(Fig.~\ref{fig:xiLumAge}).  Close pairs also tend to have formed a 
larger fraction of their stars at $z\approx 3$ than average, but 
the star forming fraction at $z<1$ of such pairs is below average 
(Fig.~\ref{fig:xiSFF}).  The objects which were forming stars at 
$z\approx 3$ have above average metallicities (Fig.~\ref{fig:xiMet}).  
These trends do not depend significantly on the mean luminosity 
of the sample, so they are approximately independent of stellar mass.  

Close pairs are dominated by galaxies in massive halos.  
Hence, our results indicate that galaxies which formed the bulk of 
their stars at high redshift are today in clusters, in which there 
is little ongoing star formation.  Since clusters formed from 
overdense regions in the early Universe, our results imply that 
cosmic star formation has moved from dense to ever less dense regions.  
This is qualitatively consistent with the findings of \citet{EDisCS06}.  
However, whereas Poggianti et al. measure instantaneous star formation 
rates in cluster galaxies identified over a wide range of redshifts 
(the SDSS at $z\sim 0$, and the ESO Distant Cluster Survey for 
$0.4\le z\le 0.8$), our method uses the spectra of the local 
galaxy population to infer the entire cosmic star formation history.  
In particular, it does not require classification of the galaxies 
into `cluster' and `field' populations, nor does it require acquisition 
of a galaxy sample which spans a wide redshift range.  
It is remarkable that these two very different methods agree.  

For similar reasons, the top left panel of Fig.~\ref{fig:xiSFF} may 
be compared with recent studies of the dependence of current star 
formation on environment \citep{Gomez03,Balogh04,SEAGal06}.  
We all find smaller star formation rates in dense regions today.
Note, however, that our analysis is not restricted to the current 
epoch---it covers 11~Gyrs in lookback time.  
Halo model interpretations of our measurements will help determine 
if the environment plays a crucial role in regulating star formation.   

\acknowledgements
RKS is supported by NASA-ATP NAG-13720 and by the NSF under 
grant AST-0520647. RJ is supported by NSF grants AST-0408698 
and PIRE-0507768, and NASA grant NNG05GG01G. 
BDP is supported by the Alexander von Humboldt Foundation, 
the Federal Ministry of Education and Research, and the 
Programme for Investment in the Future (ZIP) of the 
German Government.  

Funding for the SDSS has been provided by the 
Alfred P. Sloan Foundation, the Participating Institutions, 
the National Science Foundation, the U.S. Department of Energy, 
NASA, the Japanese Monbukagakusho, and the Max Planck Society.

\end{document}